\documentclass{article}

\usepackage[utf8x]{inputenc}
\usepackage{caption}
\usepackage[margin=0.6in]{geometry}

\usepackage{amsfonts}
\usepackage{array}
\usepackage{url}

\usepackage[american]{babel}
\usepackage{csquotes}

\usepackage{cuted}

\usepackage{graphicx}% Include figure files
\usepackage{pgf}
\usepackage{amsmath}

\newcommand\diff{\mathop{}\!\mathrm{d}}

\usepackage[
space-before-unit=true,
use-xspace=true,
list-units = bracket,
]{siunitx}

\DeclareSIUnit\sq{\ensuremath{\Box}} % for sheet resistance
\DeclareSIUnit{\percentrel}{\ensuremath{\%_{\mathrm{rel}}}}

\usepackage[inline]{enumitem}

\usepackage{booktabs}
\usepackage{multirow}
\usepackage{multicol}
\usepackage[
colorlinks=true,
linkcolor=black,
citecolor = black,
urlcolor=blue,
]{hyperref}

\usepackage{tikz}
\usetikzlibrary{calc, positioning, shapes.arrows, shadows, fadings, arrows.meta}

\definecolor{nreldarkblue}{HTML}{0B5E90}
\definecolor{nrellightblue}{HTML}{00A4E4}

\usepackage{glossaries-extra}

\setabbreviationstyle[acronym]{long-short}

\glssetcategoryattribute{acronym}{nohyperfirst}{true}

\newacronym{2t}{2T}{2-terminal}
\newacronym{3t}{3T}{3-terminal}
\newacronym{4t}{4T}{4-terminal}
\newacronym{ape}{APE}{average photon energy}
\newacronym{bc}{BC}{bottom cell}
\newacronym{cigs}{CIGS}{copper indium gallium diselenide}
\newacronym{eqe}{$Q_{c}$}{external quantum efficiency}
\newacronym{ehe}{EHE}{energy-harvesting efficiency}
\newacronym{ff}{FF}{fill factor}
\newacronym{gaas}{GaAs}{gallium arsenide}
\newacronym{gni}{GNI}{global normal irradiance}
\newacronym{ibc}{POLO-IBC}{poly-Si on oxide interdigitated back contact}
\newacronym{iv}{$I(V)$}{current-voltage}
\newacronym{jv}{$J(V)$}{current density-voltage}
\newacronym{jsc}{$J_{\mathrm{sc}}$}{short-circuit current density}
\newacronym{mpp}{MPP}{maximum power point}
\newacronym{pv}{PV}{photovoltaic}
\newacronym{psc}{PSC}{perovskite solar cell}
\newacronym{psk}{PSK}{perovskite}
\newacronym{si}{Si}{crystalline silicon}
\newacronym{sdm}{SDM}{single-diode model}
\newacronym{stc}{STC}{standard testing conditions}
\newacronym{tc}{TC}{top cell}
\newacronym{tco}{TCO}{transparent conductive oxide}

\usepackage{authblk}

\usepackage{threeparttable}

\title{Spectral effects on the energy harvesting efficiency of 2- and 4-terminal tandem photovoltaics}
\author[1]{Robert Witteck\thanks{Corresponding author: robert.witteck@nrel.gov}}
\author{John. F. Geisz}
\author{Emily. L. Warren}
\author{William. E. McMahon}
\affil[1]{National Renewable Energy Laboratory, Golden, CO, USA}
\date{September 28, 2023}
\begin{document}

\maketitle

\begin{abstract}
	In this work, we investigate how a varying spectral irradiance and top cell bandgap affect the energy yield of \gls{2t} and \gls{4t} perovskite/\!\!/silicon tandem solar cells under outdoor operating conditions. For the comparison, we first validate an optoelectronic model employing a 1-year outdoor data set for a \gls{4t} mechanical stacked \gls{gaas} on \gls{si} tandem device. We then use our verified model to simulate perovskite/\!\!/silicon tandem devices with a varying perovskite top cell bandgap for a location in Golden, Colorado, USA. We introduce a spectral binning method to efficiently reduce and improve the visualization of the \qty[number-unit-product=\text{-}]{1}{\min}-resolved environmental data while maintaining the simulation accuracy. Our findings reveal that, for a device that is current-matched under standard testing conditions, the annual spectral deviation reduces the energy harvesting efficiency by only \qty{2}{\percentrel}. When additional realistic losses for the \gls{4t} are taken into account, \gls{2t} devices are shown to have an energy-harvesting efficiency that is at parity or higher. Deviations in the top cell bandgap of more than \qty{0.1}{\electronvolt} from current matching result in a reduced energy-harvesting efficiency of more than \qty{5}{\percentrel} for the \gls{2t} tandem device.
\end{abstract}

\glsresetall

\section{Introduction} \label{sec:introduction}

Modeling of \gls{ehe} can be used to assess the relative merits of various tandem cell architectures when operated under outdoor conditions \cite{mcmahonFrameworkComparingEnergy2023}. Given the different spectral dependencies of tandem cells compared to single-junction solar cells, it is crucial to quantify their energy production under varying irradiance spectra. Here, we compare the simulated performance of \gls{2t} and \gls{4t} perovskite/\!\!/silicon tandem cells over the course of a year in Golden, Colorado (USA). Many works have already shown that \gls{2t} devices may suffer from spectral mismatch losses \cite{gotaEnergyYieldAdvantages2020, garciaSpectralBinningEnergy2018, aydinInterplayTemperatureBandgap2020, babicsOneyearOutdoorOperation2023}. In \gls{4t} devices, the top and bottom cells can operate electrically independent, each at its own \gls{mpp}. Thus, an idealized \gls{4t} device can serve as theoretical upper-limit benchmark in terms of annual energy yield performance. Consistent with the results of prior work, we find that \gls{2t} devices suffer from spectral mismatch losses. However, our results show that if the top and bottom cells are current-matched, these losses are only \qty{2}{\percentrel} when compared to an ideal \gls{4t}.

Our results include the anomalous temperature dependence for the bandgap of a \gls{psc}, which can cause the \gls{eqe} and the \gls{jsc} of a \gls{psc} to differ from those observed in traditional \gls{si} solar cells \cite{mootTemperatureCoefficientsPerovskite2021}. This can affect which sub-cell is current-limiting as the temperature changes \cite{babicsTemperatureCoefficientsPerovskite2023}. In turn, this impacts the spectral sensitivity of \gls{2t} cells and thereby their \gls{ehe}. Hence, it is reasonable to anticipate deviations in our \gls{ehe} findings for a \gls{psc}/\!\!/\gls{si} tandem compared to tandems exhibiting more conventional temperature dependencies. However, our results are quantitatively similar to prior results for more traditional tandems.

Reynolds~et~al.~\cite{reynoldsModellingPerformanceTwo2015} showed that, for micromorph \gls{si} \gls{2t} tandems, the efficiency loss for varying spectra is only \qty{2}{\percentrel} compared to that of a constant spectrum, assuming the \gls{2t} device is matched to the peak of the long-term energy distribution of the varying spectra. Futscher~et~al.~\cite{futscherEfficiencyLimitPerovskite2016} simulated an average efficiency for ideal \gls{2t} and \gls{4t} \gls{psc}/\!\!/\gls{si} tandems, resulting in an annual efficiency advantage of \qty{1.8}{\percentrel} for the \gls{4t} device for a location in Golden, Colorado. They also showed that the efficiency of a tandem cells exceeds the efficiency of a single-junction \gls{si} cell for varying incident spectra. In contrast, we simulate the \gls{ehe} of perovskite/\!\!/silicon tandems based on experimentally determined sub-cell parameters and for a minutely measured spectrum in this work.

When evaluating the performance of different tandem architectures, it is common to use an idealized \gls{4t} device as a theoretical upper-limit benchmark. However, there is an inherent design compromise for actual \gls{4t} cells. This is because the contact layer between the top cell and the bottom cell must fulfill two roles:
\begin{enumerate*}[label=(\roman*)]
	\item it must be highly conductive to allow for lateral current transport to the terminals, and
	\item it must be highly transparent for photons that will be absorbed in the bottom cell.
\end{enumerate*}
Thus, a careful \gls{ehe} analysis must consider the different optical and resistive losses that might vary with the module architecture. For instance, a monolithic series interconnection for an all-\gls{psk} tandem has different optical and resistive losses than direct deposition or stacking of a \gls{psk} \glsentrylong{tc} onto an \gls{si} \glsentrylong{bc}. Langenhorst~et~al.~\cite{langenhorstEnergyYieldAll2019} showed in a simulation study that, for a current-matched \gls{psk}/\!\!/\gls{cigs} tandem cell, the difference in energy yield between \gls{2t} and \gls{4t} devices is marginal. However, their results show that if the current-matching is poor, e.g., by using a \gls{psk} top cell with a bandgap of \qty{1.55}{\electronvolt}, the \gls{4t} device will outperform the \gls{2t} device by \qty{3.5}{\percent}. 

In this work, we calculate and compare the \gls{ehe} of \gls{2t} and \gls{4t} \gls{psk}/\!\!/\gls{si} tandem cells for varying environmental and spectral conditions of a site in Golden, Colorado. We base our modeling on measured device parameters for \gls{psc} top cells with varying bandgaps on an \gls{si} bottom cell. The environmental and spectral data covers a full year with a \qty[number-unit-product=\text{-}]{1}{\min} resolution. This enables a comprehensive temporal energy yield comparison between \gls{2t} and \gls{4t} devices, accounting for spectral and temperature effects on the sub-cell performance. We present a spectral binning method to reduce the computational effort for this comparison, which allows a fast \gls{ehe} comparison for different sites with varying spectral and environmental conditions. Besides computational benefits, the spectral binning also improves the visualization and makes differences between \gls{2t} and \gls{4t} harvesting efficiency more apparent.

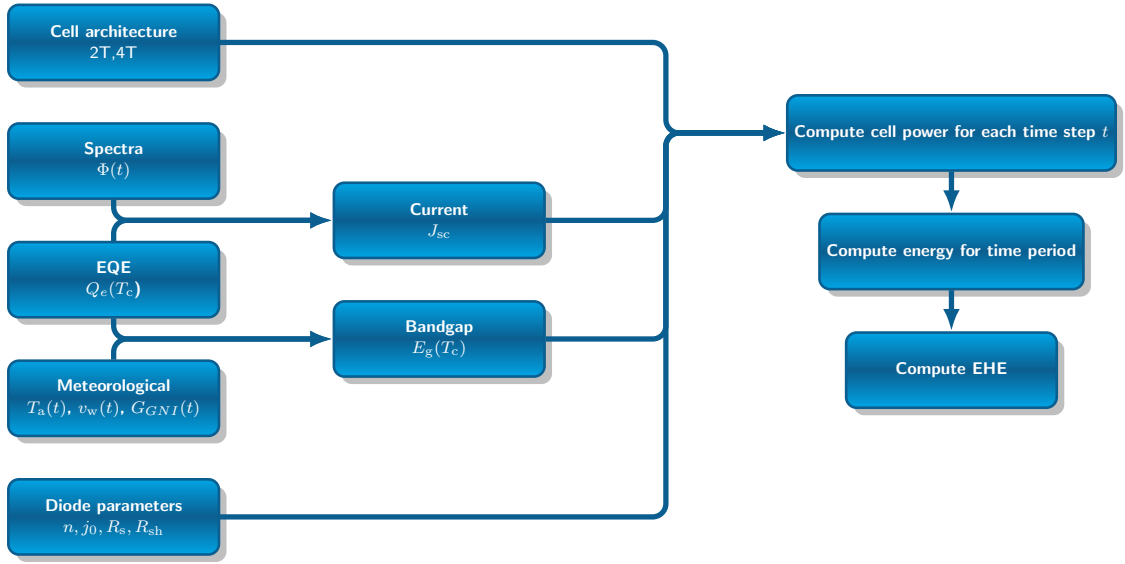
\begin{figure*}
	\centering
	%	\tikzsetnextfilename{flowchart}
	\begin{tikzpicture}[
		node distance=0.55cm and 1.5cm,
		every node/.style={
			align=center,
			anchor=north,
			scale=0.8,
		},
		arrow_style/.style={
			-{Latex[length=3mm]},
			color=nreldarkblue,
			line width=2,
			rounded corners=5pt,
		},
		vh/.style={
			arrow_style,			
			to path={
				(\tikztostart) |- (\tikztotarget)
			}
		},
		hv/.style={
			arrow_style,
			to path={
				(\tikztostart) -| (\tikztotarget)
			}
		},
		node_style/.style={
			align=center,
			%			text depth = 1.0 cm,
			draw=nreldarkblue!100!white,
			text=white,
			line width = 1,
			rounded corners,
			minimum width=3.5cm,
			minimum height=1.25cm,
			font=\fontsize{8}{8pt}\selectfont \bfseries\sffamily,
			shading=axis,
			top color=nrellightblue!100!white,
			bottom color=nrellightblue!100!white,
			middle color=nreldarkblue!100!white,
			drop shadow={shadow xshift=0.1cm, shadow yshift=-0.1cm},
		}
		]
		%		 Boxes
		% Start points
		\node[node_style] (spectra) {Spectra \\[2pt] $\Phi(t)$};
		\node[node_style, below=of spectra] (eqe) {EQE\\[2pt]$Q_{e}(T_{\mathrm{c}}$)};
		\node[node_style, below=of eqe] (meteo) {Meteorological \\[2pt]	$T_{\mathrm{a}}(t)$, $v_{\mathrm{w}}(t)$, $G_{GNI}(t)$};		
		\node[node_style, above=of spectra] (cellarch) {Cell architecture \\[2pt] {\normalfont\sffamily\footnotesize{2T,4T}}};
		\node[node_style, below=of meteo] (jv) {Diode parameters \\[2pt] $n, j_{0}, R_\mathrm{s}, R_\mathrm{sh}$};
		
		% Calc mid points
		\coordinate (midSpecEQE) at ($(spectra.south east)!0.5!(eqe.north east)$);
		\coordinate (midEQEMeteo) at ($(eqe.south east)!0.5!(meteo.north east)$);
		
		%		% inter steps
		\node[node_style, right=of midSpecEQE] (jsc) {Current \\[2pt] $J_{\mathrm{sc}}$};
		\node[node_style, right=of midEQEMeteo] (bandgap) {Bandgap \\[2pt]$E_{\mathrm{g}}(T_{\mathrm{c}})$};

		% Calc mid merge to final steps
		\node[above right=of jsc] (midMerge) {};
		%		\node[right=of jsc] (midPoint) {};
		%		\node (midMerge) at ($(jsc.east)!0.5!(midPoint.west)$) {};
		
		%		% final steps	
		\node[node_style, right=of midMerge] (compP) {Compute cell power for each time step $t$};
		\node[node_style, below=of compP] (compE) {Compute energy for time period};
		\node[node_style, below=of compE] (ehe) {Compute EHE};
		
		% Connectors
		\draw [vh] (spectra) to (jsc);
		\draw [vh] (eqe) to (jsc);
		\draw [vh] (eqe) to (bandgap);
		\draw [vh] (meteo) to (bandgap);
		\draw [arrow_style] (jsc) -| (midMerge.center) -- (compP);
		\draw [vh] (bandgap.east) -| (midMerge.center) -- (compP);
		\draw [vh] (cellarch.east) -| (midMerge.center) -- (compP);
		\draw [vh] (jv.east) -| (midMerge.center) -- (compP);
		\draw [arrow_style] (compP) to (compE);
		\draw [arrow_style] (compE) to (ehe);
		
	\end{tikzpicture}
	\caption{Modeling flow for the energy harvesting efficiency (EHE). The cell temperature $T_{c}$ inside the module is derived from the ambient temperature $T_{\mathrm{a}}$, wind speed $v_{\mathrm{w}}$, and the \glsentrylong{gni} $G_{\mathrm{GNI}}$, according to King~et~al.~\cite{kingPhotovoltaicArrayPerformance2004}.}
	\label{fig:flowchart}
\end{figure*}

\section{Modeling flow} \label{sec:modelflow}

We simulate the \gls{ehe} of different tandem cell architectures for a site in Golden, Colorado. Figure~\ref{fig:flowchart} shows our modeling flow to simulate the \gls{ehe}. The central element is \emph{PVcircuit}, an equivalent circuit-solver that facilitates the simulation of tandem solar cells with different architectures, while taking into account luminescent coupling \cite{geiszPVCircuit2022}. PVcircuit requires the spectral data, meteorological data, temperature-dependent \gls{eqe}, cell architectures, and diode parameters to simulate the \gls{ehe}. From the temperature-dependent \gls{eqe}, we obtain the temperature-dependent bandgap and \gls{jsc} of the sub-cells. We use the model from King~et~al.~\cite{kingPhotovoltaicArrayPerformance2004} to derive the cell temperature inside the \gls{pv} module from the meteorological data, i.e., from the ambient temperature, wind speed, and irradiance. We extract the diode parameters from the individual sub-cell \gls{jv} measurements. PVcircuit then simulates the power output for each time step to determine the energy yield of the cell architecture and the \gls{ehe}.

\begin{figure*}
	\centering
	\includegraphics[width=1\linewidth]{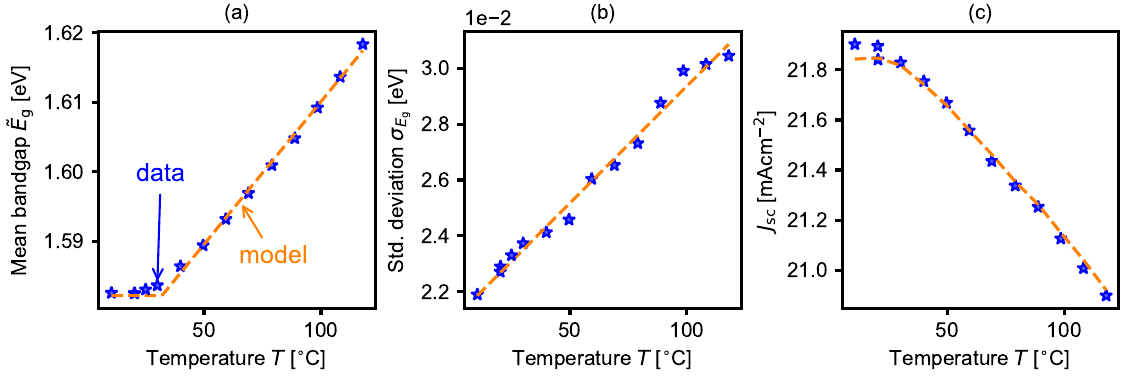}
	\caption{Temperature dependence of $\tilde{E}_{\mathrm{g}}$ and $\sigma_{E_{\mathrm{g}}}$ for the \gls{psc}. Blue symbols indicate measurements and dashed lines modeling according to Equations~\ref{eq:alphabeta} employing the parameters in Table~\ref{tab:tcoefficients}.}
	\label{fig:egsigma}
\end{figure*}

\subsection{Solar cell parameters} \label{sec:solarcellparams}

We fabricate six \glspl{psc} with varying stoichiometry and measure their \gls{jv} characteristics, transmittance, and \gls{eqe}. Detailed information on the cell processing procedures is available in Tong~et~al.~\cite{tongCarrierLifetimesMs2019} and Kim~et~al.~\cite{kimBimolecularAdditivesImprove2019}. The stoichiometric variation results in \glspl{psk} with varying bandgaps $E_{\mathrm{g,25}}$ of \qtylist{1.58; 1.63; 1.68; 1.7; 1.75; 1.8}{\electronvolt}, as determined from \gls{eqe} measurements at \qty{25}{\degreeCelsius} \cite{rauEfficiencyPotentialPhotovoltaic2017}. A \gls{sdm} is used to extract the diode parameters from the measured \gls{jv} characteristics. These \glspl{psc} serve as potential top cells.
For the \gls{si} bottom cell, we measure the \gls{jv} characteristics and \gls{eqe} of an \gls{ibc} Si cell and similarly fit the \gls{jv} data with the \gls{sdm}. We integrate the AM1.5G convolved \gls{eqe} to determine the \gls{jsc} of the bottom cell in a tandem solar cell, accounting for optical transmission through the various top cells. Using a \gls{psk} top cell with a bandgap of \qty{1.70}{\electronvolt} results in a current-matched \gls{2t} \gls{psk}/\!\!/\gls{si} tandem device for our experimental data. 

\subsection{Bandgap extraction and bandgap temperature dependence} \label{sec:bandgap_params}

We measure the temperature-dependent \gls{eqe} of the \gls{si} as well as the \gls{psk} solar cell with the bandgap of \qty{1.58}{\electronvolt} and extract the bandgap for each temperature similar to the method of Rau~et~al.~\cite{rauEfficiencyPotentialPhotovoltaic2017}. We fit the first derivative of the \gls{eqe} with a Gaussian function and extract the mean bandgap $\tilde{E}_{\mathrm{g}}$ and the standard deviation $\sigma_{E_{\mathrm{g}}}$. Figure~\ref{fig:egsigma}\,(a) shows $\tilde{E}_{\mathrm{g}}$, and Figure~\ref{fig:egsigma}\,(b) shows $\sigma_{E_{\mathrm{g}}}$ for the \gls{psc}. For the temperature dependence of $\tilde{E}_{\mathrm{g}}$ of the \gls{psc}, we assume a piecewise linear function, with a constant bandgap for temperatures below \qty{32}{\degreeCelsius} and a linearly increasing bandgap for higher temperatures. This is in agreement with other bandgap temperature dependencies reported in the literature \cite{mootTemperatureCoefficientsPerovskite2021, liHighlyEfficientPin2023}. For $\sigma_{E_{\mathrm{g}}}$ of the \gls{psc} as well as the $\tilde{E}_{\mathrm{g}}$ and $\sigma_{E_{\mathrm{g}}}$ for the \gls{si} cell, we assume a linear dependence for all temperatures
\begin{align}
	\tilde{E}_{\mathrm{g}} = \alpha T + \beta && \sigma_{E_{\mathrm{g}}} =  a T + b,
	\label{eq:alphabeta}
\end{align}
where $T$ is the temperature and $\alpha$, $\beta$, $a$, and $b$ are material-specific coefficients, as summarized in Table~\ref{tab:tcoefficients}.

\begin{table}
	\begin{threeparttable}
		\centering
		\sisetup{
			table-alignment-mode = format,
			table-number-alignment = center,
			table-format= +1.2e-1,
			round-mode = places,
			round-precision = 2,
			scientific-notation = true,
			table-column-width = \linewidth/6,
			%		table-sign-mantissa,
		}
		\renewcommand{\arraystretch}{1.2}%
		\caption{Temperature coefficients for $\tilde{E}_{\mathrm{g}}$ and $\sigma_{E_{\mathrm{g}}}$}
		\begin{tabular*}{\linewidth}{S S S S S}
			\toprule
			{Cell} & {$\alpha$ [\unit{\electronvolt\per\degreeCelsius}]} & {$\beta$ [\unit{\electronvolt}]} & {$a$ [\unit{\electronvolt\per\degreeCelsius}]} & {$b$ [\unit{\electronvolt}]}\\
			\hline
			{\gls{psc} \tnote{a}} & 4.10879270e-04 & 1.56900992e+00 & 8.36429025e-05 & 2.09940616e-02 \\
			{\gls{si}}  & -6.26701229e-04 & 1.17062177e+00 & 0.00027643 & 0.05863486 \\
			\bottomrule
		\end{tabular*}
		\label{tab:tcoefficients}
		\begin{tablenotes}
			\item[a] $\tilde{E}_{\mathrm{g}}(\alpha, \beta)$ if $T>$\qty[scientific-notation = false, round-precision = 0]{32}{\degreeCelsius} and \qty{1.58}{\electronvolt} otherwise.
		\end{tablenotes}
	\end{threeparttable}
\end{table}

Employing Equation~\ref{eq:alphabeta} with the parameters in Table~\ref{tab:tcoefficients}, we can determine $\tilde{E}_{\mathrm{g}}$ and $\sigma_{E_{\mathrm{g}}}$ for a given temperature and derive the corresponding \gls{eqe}. Figure~\ref{fig:egsigma}\,(c) demonstrates the applications of this approach for the \gls{psc}. The symbols indicate the \gls{jsc} extracted from the measured \gls{jv} data, and the dashed line indicates the \gls{jsc} resulting from the integration of the AM1.5G convoluted and temperature-corrected \gls{eqe}.
%The red symbols for the second ordinate axis indicate the difference in \gls{jsc} between measurement and derived values. Comparing both methods results in a deviation of less than \qty{0.1}{\milli\ampere\per\square\centi\meter}, which demonstrates the applicability of the method. 
The advantage of using the \gls{eqe} to determine the \gls{jsc} is that, besides considering the temperature dependence, it also accounts for the mutual influence of top cell and bottom cell \gls{eqe}. In the energy yield modeling, we assume that this temperature dependence is the same for all \glspl{psc} of varying bandgaps.

\subsection{Diode model} \label{sec:diode_model}

We model the output voltage $V_{\mathrm{cell}}$ and current density $J_{\mathrm{cell}}$ of the individual top and bottom solar cells using PVcircuit \cite{geiszGeneralizedOptoelectronicModel2015, geiszCharacterizationMultiterminalTandem2021}. PVcircuit employs a diode model with
\begin{equation*}
	V_{\mathrm{cell}} = V_{\mathrm{diode}} - JR_{\mathrm{s}}
%	\label{eq:v_cell}
\end{equation*}
and
\begin{equation*}
	J_{\mathrm{cell}} = J_{\mathrm{sc}} - J_{\mathrm{rec}} - G_{\mathrm{sh}} V_{\mathrm{diode}},
%	\label{eq:j_cell}
\end{equation*}
where $G_{\mathrm{sh}}$ is the shunt conductance, $V_{\mathrm{diode}}$ is the voltage across the diode, $J_{\mathrm{sc}}$ is the cell's short-circuit current density, and $J_{\mathrm{rec}}$ is the cell's recombination current density. The latter is calculated by
\begin{equation*}
	J_{\mathrm{rec}}  \left(V_{\mathrm{diode}}\right) = \sum_{j=1}^{N_\mathrm{d}} J_{0j} \left[ \exp \left( \frac{q V_{\mathrm{diode}}}{n_{0j}kT} \right) - 1 \right],
%	 \label{eq:jrec}	
\end{equation*}
where $J_{0j}$ is the diode saturation current density, $n_{0j}$ is the ideality factor of the $j^{\mathrm{th}}$ diode, $q$ is the elementary charge, $k$ is Boltzmann's constant, $T$ is the temperature, and $N_\mathrm{d}$ is the number of parallel diodes. For the sake of simplicity, we use a single-diode model with $N_\mathrm{d}=1$ throughout this work. Note that in principle, PVcircuit has the capability to simulate multiple diodes.

The model assumes that the $J_{0j}$ scales by a constant $A_{j}$, with the detailed-balance saturation current density $J_{\mathrm{db},i}$ according to
\begin{equation*}
	J_{0j} \left(T\right) = A_{j} \left[J_{\mathrm{db},i}\right]^\frac{1}{n_{0j}}.
	\label{eq:j0i}
\end{equation*} 
$J_{\mathrm{db},i}$ depends on the mean bandgap $\tilde{E}_{\mathrm{g},i}$ and the standard deviation $\sigma_{E_{\mathrm{g}},i}$ of the $i^{\mathrm{th}}$ solar cell \cite{rauRadiativeEfficiencyLimits2004}
\begin{strip}
	\hrulefill
	% \rule{0.9\linewidth}{0.33pt}
	\begin{equation*}
		J_{\mathrm{db},i} \left(\tilde{E}_{\mathrm{g},i}, T\right) = \frac{2\pi q k^3 T^3} {h^3 c^2} \Bigg[ \\
		\frac{\tilde{E}_{\mathrm{g},i}^2} {k^2 T^2}
		+ 2\frac{\tilde{E}_{\mathrm{g},i}} {k T}
		+ 2 \\
		- 2 \frac{\sigma_{E_{\mathrm{g}}}^2 \tilde{E}_{\mathrm{g},i}} {k^3 T^3}
		- \frac{\sigma_{E_{\mathrm{g}}}^2} {k^2 T^2}
		+ \frac{\sigma_{E_{\mathrm{g}}}^4} {k^4 T^4}  \\
		\Bigg] \exp \left(-\frac{\tilde{E}_{\mathrm{g},i}} {k T} \right) ,
%		 \label{eq:jdb}
	\end{equation*}
	% \hfill\rule{0.9\linewidth}{0.33pt}
	\hrulefill
\end{strip}

where $c$ is the speed of light. 

We obtain $A_{j}$, $n_{0j}$, and $J_{\mathrm{sc}}$ from fitting the diode equation to \gls{jv} measurements of individual solar cells at \gls{stc}. PVcircuit also accounts for luminescent coupling between the sub-cells. Here, we follow the approach of Geisz~et~al.~\cite{geiszGeneralizedOptoelectronicModel2015}, employing a $\beta$ of 16 for the bottom cell.

\subsection{Meteorological data} \label{sec:environmental}

We use the environmental data measured at the NREL Solar Radiation Research Laboratory (SRRL) in Golden, Colorado, to simulate the energy yield of the tandem devices \cite{stoffelNRELSolarRadiation1981}. The data set spans a period of one year (November 1, 2021 until October 31, 2022) and has a \qty[number-unit-product=\text{-}]{1}{\min} resolution. For the spectral data, we use the measurements from the two-axis tracked EKO Instruments WISER MS-712 spectroradiometer, denoted $\Phi_{\mathrm{GNI}}$. We calculate the resulting \gls{gni} as
\begin{equation*}	
	G_{\mathrm{GNI}} = \int_{290}^{1650} \Phi_{\mathrm{GNI}} \diff\lambda,
%	\label{eq:ggni}
\end{equation*}
where $\lambda$ is the wavelength.

To derive the cell temperature inside the module, we employ the model from King~et~al.~\cite{kingPhotovoltaicArrayPerformance2004}, which takes the irradiance, wind speed, and ambient temperature into account. Data affected by instrument maintenance schedules and spectra with
% more than \qty{10}{\percent}
negative values are discarded.

\subsection{Model verification}

\begin{figure}[t]
	\centering
	\includegraphics[width=0.87\linewidth]{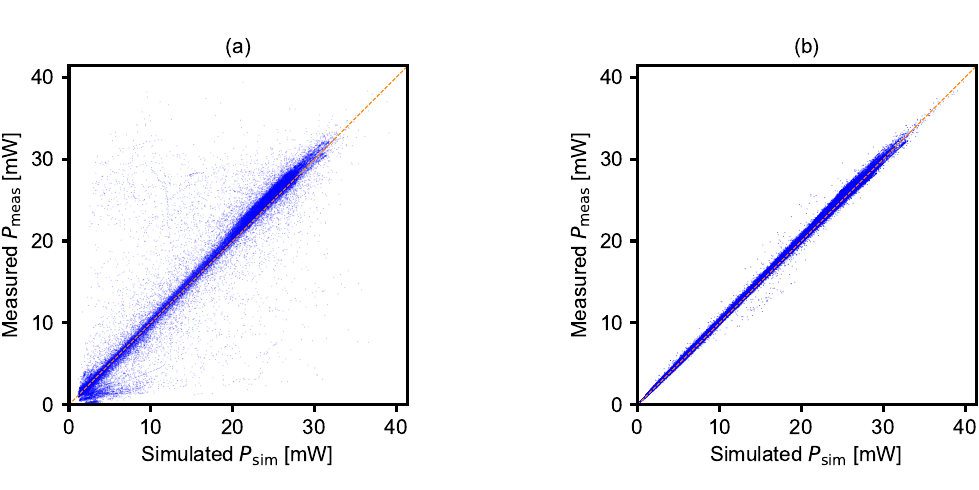}
	\caption{Measured vs. simulated module power for the \gls{4t} GaAs/\!\!/\gls{si} tandem when (a) the \gls{jsc} from the integrated \gls{eqe} convoluted spectra is used as an input parameter for PVcircuit and (b) the measured current is used as input parameters.}
	\label{fig:pvsp}
\end{figure}

We compare the power output simulated with PVcircuit to the measured power output of a GaAs/\!\!/\gls{si} \gls{4t} tandem solar module \cite{whiteheadOptimizationFourTerminal2021} to verify our model. The module was monitored for one year at the location in Golden, Colorado. We chose a \gls{gaas}-on-\gls{si} \gls{4t} tandem cell for the model verification because both materials are more resilient to outdoor degradation and operate more stable than e.g., \gls{psk} \cite{emeryEncapsulationOutdoorTesting2022, silvermanBriefDailyPerformance2023}. Hence, the associated cell parameters needed for numerical modeling did not change significantly over the course of the year. A \gls{iv} curve tracer measures the \gls{iv} characteristics of the top and bottom cell every \qty{2}{\min}. For further details on the tandem module and the measurements, we refer to Springer~et~al.~\cite{springer2023}.

Following the model flow in Figure~\ref{fig:flowchart}, we simulate the module power of the tandem device using measured site spectra and meteorological data. Figure~\ref{fig:pvsp}\,(a) shows an identity plot comparing the measured and simulated module power. The orange dashed line is the identity line (where the measured module power is equal to the simulated module power). The majority of the data points align along this identity line. We use the Pearson correlation coefficient $R$ to quantity the correlation between measurement and simulation. For the data in Figure~\ref{fig:pvsp}\,(a), we obtain an $R$-value of 0.9697, where an $R$-value of 1.0 means an exact match of measurement and simulation. This excellent agreement demonstrates the ability of PVcircuit to model the outdoor data with sufficient accuracy. 

The variations between the measured and simulated module power stem primarily from discrepancies in the \gls{jsc} values. Figure~\ref{fig:pvsp}\,(b) shows the measured and simulated module power when using the measured \gls{jsc}, instead of the \gls{jsc} from integrated \gls{eqe} convoluted spectra, as input parameters for PVcircuit. This reduces the number of data points deviating from the identity line and improves the $R$-value to 0.9993. We ascribe the differences in \gls{jsc} to the different locations of the spectroradiometer and the solar module test site, as well as asynchronous measurement times. In addition, the measured \gls{jsc} accounts for a small degradation that we have observed over the year (see \cite{springer2023}), which is neglected in the \gls{jsc} determined from the integrated \gls{eqe}.

\subsection{Energy harvesting efficiency of \gls{psc}/\!\!/\gls{si} tandems} \label{sec:ehe}

After verifying PVcircuit for energy yield modeling, we use the environmental data from Section~\ref{sec:environmental} to model the annual energy output $E_{\mathrm{out}}$ of a \gls{2t} and a \gls{4t} \gls{psk}/\!\!/\gls{si} tandem solar module. The annual energy produced by the solar module is calculated as
\begin{equation*}
	E_\mathrm{out} = \sum\nolimits_{t} P_{\mathrm{out}} \; t,
\end{equation*}
where $P_{\mathrm{out}}$ is the module power output for each time step $t$.
The \gls{ehe} allows us to quantify the efficiency of energy harvesting by considering the temporal and spectral aspects of the solar module's performance, and it relates the produced annual energy to the incident solar energy $E_{\mathrm{GNI}}$ as follows
\begin{equation}
	\eta_{\mathrm{ehe}}
	= \frac{E_\mathrm{out} } {E_\mathrm{GNI}}
	= \frac{E_\mathrm{out} } {\sum\nolimits_{t} E_{\mathrm{GNI},t}}
	= \frac{\sum\nolimits_{t} P_{\mathrm{out}} \; t} {\sum\nolimits_{t} G_{\mathrm{GNI}}  \; t}.
	%		= \frac{\sum\nolimits_{t} G_{\mathrm{GNI}} t} {\sum\nolimits_{t} \int \Phi \diff\lambda \; t},
	\label{eq:eta_ehe}
\end{equation}

\section{Energy yield simulations}

\subsection{Spectral binning} \label{sec:binning}

\begin{figure*}
	\centering
	\includegraphics[width=1.\textwidth]{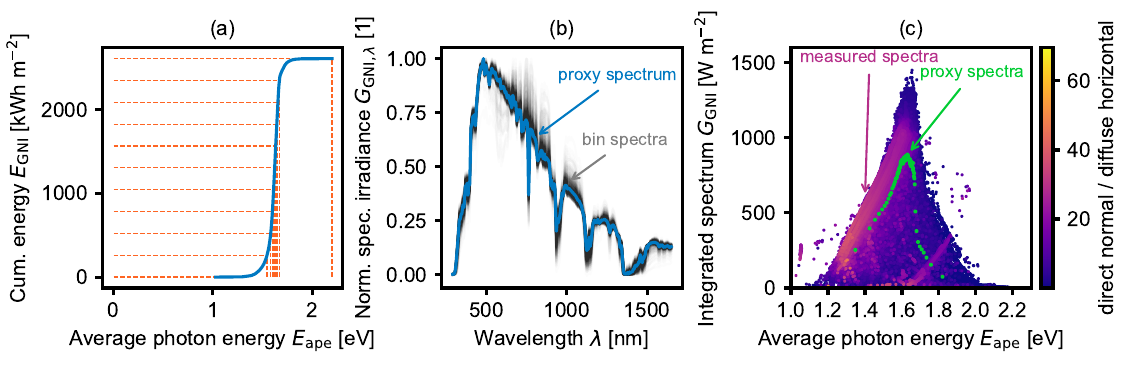}
	\caption{(a) Spectral binning by equal energy into 10 bins. The orange dashed lines show the bin edges, and the blue line shows the corresponding  $E_{\mathrm{GNI},k}$. (b) The resulting proxy spectra of the fifth bin with a mean \gls{ape} of \qty{1.63}{\electronvolt}, shown in blue. The underlying gray lines show every $20^{\mathrm{th}}$ spectrum in the bin. (c) The integrated irradiance for all spectra, colored by the direct-to-diffuse irradiance ratio. The superimposed green dots indicate the integrated irradiance of 100 proxy spectra.}
	\label{fig:binning}
\end{figure*}

Simulating the \gls{ehe} for the full year using \qty[number-unit-product=\text{-}]{1}{\min} time intervals is computationally expensive. Thus, we reduce the amount of data by employing a spectral binning method that bins the spectra into sets of similar spectra. Each bin of spectra can then be represented with a single "proxy" spectrum. The method follows the approach presented by Garcia~et~al.~\cite{garciaSpectralBinningEnergy2018}, but instead of the equivalent photocurrent ratio, we use the \gls{ape}. The \gls{ape} of a solar spectrum is given by \cite{jardineInfluenceSpectralEffects2002}
\begin{equation*}
	E_{\mathrm{ape}} = \frac{hc} {q} \frac{\int \Phi_{\mathrm{GNI}} \; \mathrm{d}\lambda} {\int \Phi_{\mathrm{GNI}} \lambda \diff\lambda},
\end{equation*}
where $h$ is Planck's constant.

We assign each spectrum an \gls{ape} and incident energy
\begin{equation*}
	E_{\mathrm{GNI}} = G_{\mathrm{GNI}} \; t.
\end{equation*}
Subsequently, we sort the spectra according to their \glspl{ape} and calculate the cumulative sum of the energy for all spectra. We divide the maximum of the cumulative sum into $N_{Sbins}$ equal energy bins. Figure~\ref{fig:binning}\,(a) shows this for 10 bins. The orange dashed lines indicate the bin edges segmenting the cumulative sum of the energy, thus conserving the energy for each bin.

The $k^{\mathrm{th}}$ bin out of the $k=1\dots N_{Sbins}$ bins contains $n_k$ spectra. We assign each of the $l=1\dots n_k$ spectra $\Phi_{kl}$ in a bin a time $t_{kl}$ and temperature $T_{kl}$. %, and irradiance $G_{\mathrm{GNI},i,j}$.%
Thus, the total time for each bin $t_k$ is given as
\begin{equation*}
	t_k = \sum_{l = 1}^{n_k} t_{kl}.
\end{equation*}
We can then assign each bin an average proxy spectrum by taking the mean of all spectra in this bin
\begin{equation*}
	\Phi^{p}_{k} = \frac{1}{t_k}\sum_{l = 1}^{n_k} \Phi_{kl} t_{kl}. %\qquad \forall \quad 1 < i \leq N_{Sbins}.
\end{equation*} 
Figure~\ref{fig:binning}\,(b) illustrates this for the proxy spectra of the fifth bin in blue, with a mean \gls{ape} of \qty{1.64}{\electronvolt}. The underlying gray lines show every $20^{\mathrm{th}}$ spectrum in the bin. The original spectra are plotted in grey with some transparency, so overlapping spectra become darker in the graph. It is important to note that the proxy spectrum in blue represents the mean of all gray spectra, since the mean mathematically conserves the energy.

Figure~\ref{fig:binning}\,(c) shows the integrated irradiance for all spectra colored by the direct/diffuse irradiance ratio. The integrated irradiance of the 100 proxy spectra is indicated by the superimposed green dots. The direct/diffuse ratio reveals that clear sky days with a high fraction of direct light are "red-rich" and have a lower \gls{ape}. More diffuse light is "blue-rich" due to scattering of short wavelength photons by clouds, and has a higher \gls{ape}. However, for the blue-shifted tail of the distribution, the higher cloud coverage also significantly reduces the irradiance. Consequently, these spectra provide light with shorter wavelengths but less incident energy over the course of the year. As a result, \qty{90}{\percent} of the energy of all spectra are in the range between \qtyrange{1.50}{1.70}{\electronvolt}, consistent with Figure~\ref{fig:binning}\,(a). Besides reducing the computational effort, the binning also helps visualize the relevant spectra that have the biggest impact on the energy yield.

\begin{figure*}
	\centering	
	\includegraphics[width=0.87\linewidth]{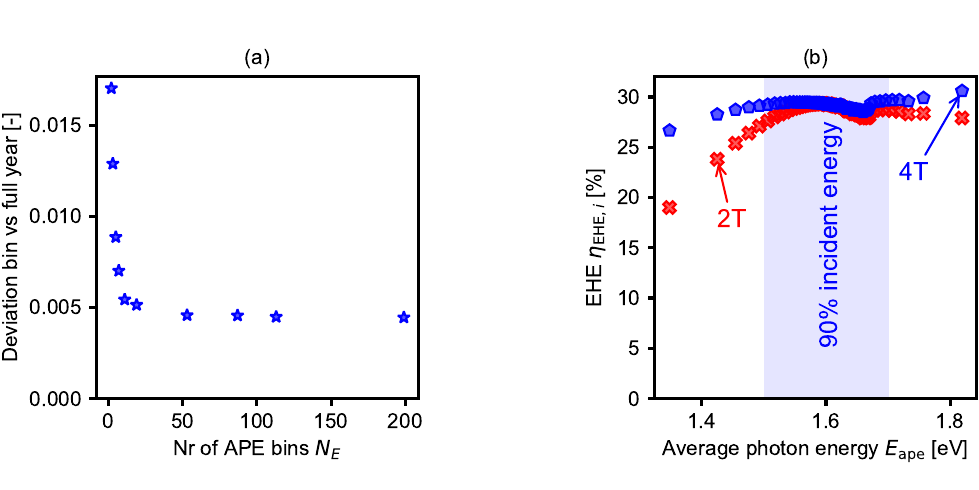}
	\caption{(a) Deviation in energy yield for the \gls{2t} device, when comparing the full-year energy yield with the equal energy binning approach using five temperature bins per spectral bin. (b) Comparison of the \glsentrylong{ehe} for \gls{2t} and \gls{4t} devices for 100 proxy spectra. The blue area highlights the range that accounts for \qty{90}{\percent} of the energy of all incident spectra.}
	\label{fig:2t4t_bin-fullyear}
\end{figure*}

In addition to the irradiance, the cell temperature inside the module is crucial for the outdoor performance of the tandem. We bin the temperatures of each spectral bin, denoted as $T_{kl}$, into $N_{Tbins}$ cell temperature bins with equal bin width $\Delta T$. We then simulate the power output for each of the $N_{Tbins}$ temperature bins for a given proxy spectra to calculate the \gls{ehe} for this spectral bin as
\begin{equation}
	\eta_{\mathrm{ehe},k} = \frac{\sum_{m = 1}^{N_{Tbins}} P_{\mathrm{out},km} \; t_{km}}{\int \Phi_{k}^{p} \diff\lambda \; t_{k}}.
	\label{eq:ehei}
\end{equation}
For each temperature bin, we employ its median temperature value. The \gls{ehe} for all proxy spectra is then given by
\begin{equation*}
	\eta_{\mathrm{ehe}} = \frac{\sum_{k = 1}^{N_{Sbins}} \sum_{m = 1}^{N_{Tbins}} P_{\mathrm{out},km} \; t_{km}} {\sum_{k = 1}^{N_{Sbins}} \int \Phi_{k}^{p} \diff\lambda \; t_{k}}.
%	\label{eq:ehe}
\end{equation*}
With an increasing number of spectral and temperature bins, this will approximate the expression in Equation~\ref{eq:eta_ehe}.

\subsection{Comparing spectral binning and full-year EHE simulations}

We compare the equal energy binning approach with full-year simulations by simulating a \gls{2t} and a \gls{4t} device. For a fair comparison between \gls{2t} and \gls{4t} devices, we choose a \gls{psk} top cell with a bandgap of \qty{1.70}{\electronvolt}, as this results in a current-matched device for \gls{stc} measurements with AM1.5G spectrum. For the full-year simulation, we simulate the power output of the \gls{2t} tandem device for each time step and calculate the annual energy yield. For the binning approach, we vary the number of bins $N_{Sbins}$, simulate the power output of the \gls{2t} tandem device, and calculate the annual energy considering the total time of each bin.%

Figure~\ref{fig:2t4t_bin-fullyear}\,(a) shows the deviation in energy yield for the \gls{2t} device when comparing the full-year energy yield simulation with the binning approach. For each of the $N_{Sbins}$ variations, we use five temperature bins. Employing more than 50 bins reduces the deviation between the full-year and binning approach to less than \qty{0.5}{\percent}. This is consistent with other methods of agglomerating spectral data using other binning methods \cite{garciaSpectralBinningEnergy2018} or machine learning methods \cite{ripaldaLocationSpecificSpectralThermal2020}. Hence, this approach allows for a significant reduction in computational effort when comparing the \gls{ehe} for different top-cell bandgaps combined with an \gls{si} bottom cell in \gls{2t} and \gls{4t} devices, while still ensuring accurate simulation results. For instance, on a conventional laptop, a full-year simulation takes several days, while a 50-bin simulation takes only a few minutes.

\subsection{Comparing the EHE of 2T and 4T tandem cells}

In this section, we compute the \gls{ehe} for tandem cells comprising of a \qty[number-unit-product=\text{-}]{1.70}{\electronvolt} \gls{psc} top cell and an \gls{si} bottom cell over the course of one year in Golden, Colorado. We choose the top cell with a bandgap of \qty{1.70}{\electronvolt}, as this results in a current-matched tandem under the AM1.5G spectrum when employing an \gls{si} bottom cell. This enables a comparison of \gls{2t} and \gls{4t} cells, where both operate close to their optimum. We initially neglected resistive and optical losses to focus on losses due to spectral variation. These will be reintroduced later for a more complete comparison.

As described in the previous section, we use a spectral-binning algorithm to reduce the full set of spectra to a much smaller set in which proxy spectra represent sets of spectra with similar \gls{ape} values. The cell power and \gls{ehe} are computed for each proxy spectrum, as well as for the entire year, by considering the time for each bin.

Figure~\ref{fig:2t4t_bin-fullyear}\,(b) shows the computed efficiencies for 100 proxy spectra according to Equation~\ref{eq:ehei}. The \glspl{ape} of the proxy spectra range from \qtyrange{1.35}{1.82}{\electronvolt}. 
The \gls{2t} device is nearly current-matched for a spectra with an \gls{ape} of \qty{1.60}{\electronvolt}, so for spectra near this energy, the \gls{2t} and \gls{4t} efficiencies are nearly identical. Consequently, both sub-cells operate in close proximity to their respective \glspl{mpp}. As the incident proxy spectra become more red- or blue-rich, the \gls{2t} efficiency becomes lower, because the sub-cells are no longer current-matched. However, \qty{90}{\percent} of the incident energy is provided by spectra with \gls{ape} values between \qty{1.50}{\electronvolt} and \qty{1.70}{\electronvolt} (shaded blue region). In this region, the \gls{2t} efficiency is close to the \gls{4t} efficiency. As a result, the annual \gls{ehe} of the \gls{2t} device is only \qty{2}{\percentrel} lower than the \gls{ehe} of the \gls{4t} device. This result serves as the baseline for comparison in Table~\ref{tab:4tloss}, where various scenarios of \gls{4t}-related optical and resistive losses are examined. The upshot is that despite encountering spectral variations during outdoor operation, the \gls{ehe} of an \gls{stc} current-matched \gls{2t} is only slightly smaller than that of a \gls{4t} device. This is consistent with the results for more traditional tandems \cite{reynoldsModellingPerformanceTwo2015}, regardless of the anomalous temperature behavior of the \gls{psc}. One possible explanation is that operating the cell at the \gls{mpp} makes the spectral and temperature effects marginal due to \gls{ff} compensation \cite{wanlassRigorousAnalysisSeriesConnected2004,mcmahonFillFactorProbe2008, reynoldsModellingPerformanceTwo2015,blomEnergyLossAnalysis2023}.

Furthermore, the architectural differences between \gls{2t} and \gls{4t} devices generally lead to distinct optical and resistive losses. The most significant difference is that a \gls{2t} device typically employs a thin tunnel or recombination junction in between the sub-cells, which creates low optical and resistive losses \cite{bastianiRecombinationJunctionsEfficient2020,zhangEfficientInterconnectingLayers2022}. In contrast, a \gls{4t} device requires an additional transport layer to facilitate lateral current transport between the sub-cells. Typically, this is accomplished by implementing a \gls{tco} layer between the sub-cells. Thus, there is a trade-off between optical and resistive losses associated with this \gls{tco} layer. Assuming the \gls{tco} layer contributes an additional series resistance of \qty[round-mode=places,round-precision=2]{1.86577}{\ohm\square\centi\meter} results in parity in the annual \gls{ehe} between \gls{2t} and \gls{4t} devices. Similarly, optical losses of \qty{5}{\percent} for the bottom cell current, due to free carrier absorption in the \gls{tco} layer, would equalize the energy yield. Table~\ref{tab:4tloss} summarizes the results for both assumptions. Exceeding any of these inherent and unavoidable losses in the \gls{4t} device will result in the \gls{2t} device having a greater \gls{ehe} than the \gls{4t} device.

\begin{table}
	\centering
	\sisetup{
		table-number-alignment = center,
		table-alignment-mode = format,
		table-format= 3.2,		
		table-column-width = \linewidth/5,
	}
	\renewcommand{\arraystretch}{1.2}%
	\caption{Considered electrical and optical losses when integrating a \gls{4t} into a solar module.}
%	\begin{tabular*}{\linewidth}{@{\extracolsep{\fill}}S S S S@{}}
	\begin{tabular*}{\linewidth}{S S S S}
		\toprule
		{Scenario} & {Rs} & {TCO loss 4T} & {2T/4T energy yield} \\
		& [\unit{\ohm\square\centi\meter}] & [\unit{\percent}] & [\unit{\percent}] \\
		\hline
		1 & 0 & 0 & 98\\
		2 & 1.86 & 0 & 100\\
		3 & 0 & 5 & 100\\
		\bottomrule
	\end{tabular*}
	\label{tab:4tloss}
\end{table}

Usually, both losses are intertwined and occur simultaneously. As an example, for our \gls{tco} with a sheet resistance of \qty[per-mode=symbol]{24}{\ohm\per\sq}, we obtain an additional series resistance contribution of \qty{1.82}{\ohm\square\centi\meter} and optical losses of \qty{2.2}{\percent} for the bottom cell, which results in a higher \gls{ehe} for the \gls{2t} device than for the \gls{4t} device. The exact losses will be situational and unique to specific devices. However, it can be beneficial to initially conduct a comparison with optical and resistive losses neglected, as this establishes a baseline for comprehending the significance of the different losses involved.

\begin{figure}
	\centering
	\includegraphics[width=0.87\linewidth]{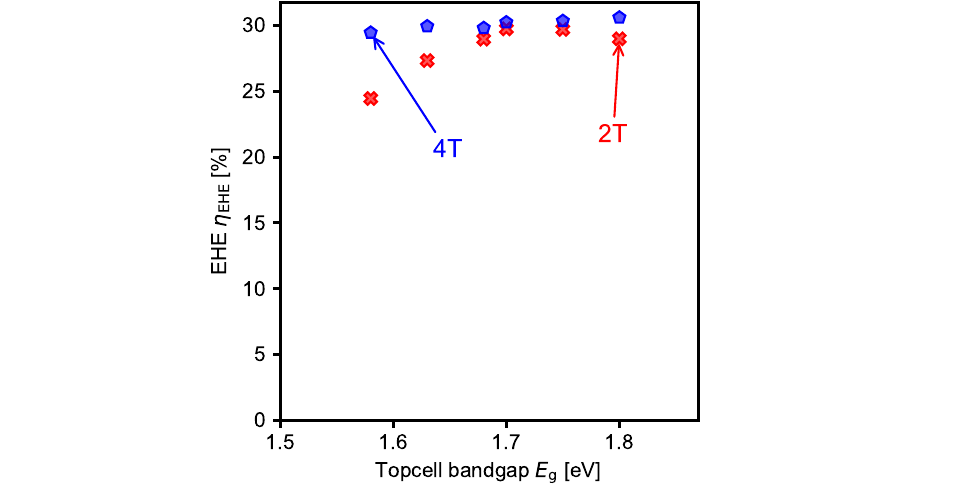}
	\caption{Annual \gls{ehe} for a perovskite/\!\!/silicon tandem cell with varying \glsentrylong{tc} bandgap and \gls{si} \glsentrylong{bc}.}
	\label{fig:2t4t_bg}
\end{figure}

Given that current-matching is crucial for achieving peak \gls{2t} device performance, the choice of sub-cell bandgaps significantly impacts the \gls{2t} versus \gls{4t} \gls{ehe} comparison. Figure~\ref{fig:2t4t_bg} shows the simulated annual \gls{ehe} for a 1-year operation in Golden. We simulate tandem cells with varying bandgaps for the \gls{psk} top cell featuring an \gls{si} bottom cell. In the calculations, we assume no additional resistive or optical losses for the \gls{4t} device, resulting in the previously mentioned \gls{ehe} advantage of \qty{2}{\percent} for the current-matched device with a \qty[number-unit-product=\text{-}]{1.70}{\electronvolt} bandgap for the top cell. Considering a non-optimal bandgap for the top cell of \qty{1.58}{\electronvolt} results in a \qty{17}{\percentrel} lower annual \gls{ehe} for the \gls{2t} device due to the higher current mismatch. Such high losses will be challenging to compensate for, even when considering additional losses due to the \gls{tco} layer which is necessary for the module integration of the \gls{4t} device. Hence, it is imperative to consider a well-current-matched device for the \gls{2t} tandem. Small variations in the bandgap of \qty{0.05}{\electronvolt} only have a marginal impact on the annual \gls{ehe}.
However, it is important to acknowledge that wide-bandgap materials may exhibit additional challenges that can adversely impact the properties of the sub-cells \cite{ganoseDefectChallengeWidebandgap2022, yangDefectEngineeringWidebandgap2022, laiHighPerformanceFlexibleAllPerovskite2022}.

Finally, the test site location in Golden has an average \gls{ape} of \qty{1.61}{\electronvolt}, which is close to the \gls{ape} of \qty{1.62}{\electronvolt} of the AM1.5G spectrum in the same wavelength range. Thus, the ideal combination of top cell and bottom cell bandgaps for a current-matched device may be different for locations that vary significantly from this value.

\section{Summary and outlook}

In this work, we provide a framework to perform energy yield simulations for \gls{2t} and \gls{4t} tandem solar cells considering varying outdoor spectra and top cell bandgaps. We verify our model by comparing the measured and simulated power output of a GaAs/\!\!/\gls{si} tandem that was deployed outdoors for one year. Employing the measured device current as input parameters results in an excellent agreement between measurement and simulation, with a Pearson correlation coefficient of 0.9993. 

We introduce a new spectral binning method that uses the \glsentrylong{ape} to agglomerate the spectral data. Besides reducing the computational effort, the binning approach also improves the data visualization to provide a better understanding of spectral effects on the tandem performance. Varying the number of bins and comparing it to the full-year simulation, %with 265,667 data points
we find that employing more than 50 bins reduces the deviation between full-year and binning approach to less than \qty{0.5}{\percent}. Thus, the binning method significantly reduces the computational effort for energy yield simulations while providing sufficient accuracy to compare the energy yield of different tandem architectures. One advantage of using the \gls{ape} for the spectral binning is that this method is independent of any device parameter assumptions. Compared to other binning techniques, such as machine learning, our method offers a simple and computationally fast alternative.

A comparison of the \gls{ehe} between a current-matched device with \gls{2t} and \gls{4t} architecture shows only a \qty{2}{\percent} penalty for the \gls{2t} device. Considering additional resistive losses of \qty[round-mode=places,round-precision=2]{1.86577}{\ohm\square\centi\meter} or optical losses of \qty{5}{\percent} due to a necessary \gls{tco} layer for the module integration of a \gls{4t} device results in parity for both architectures in terms of energy yield. A first calculation with our in-house \gls{tco} indicates that this can be a realistic assumption for the required \gls{tco} layer. A detailed investigation of the optical and electrical properties of the \gls{tco} layer on various tandem architectures is planned for future work. 

The variation of the top cell bandgap shows that it is essential to have a current-matched device to achieve similar \gls{ehe} for \gls{2t} and \gls{4t} devices. A device that is not current-matched, e.g., with a bandgap of \qty{1.58}{\electronvolt}, results in a \qty{17}{\percentrel} lower \gls{ehe} for the \gls{2t} device compared to the \gls{4t} device. In a future study, we plan to investigate whether this is also true for other sites, which may show a higher deviation of the spectra from the AM1.5G spectrum.

% Acknowledgements
\medskip
\textbf{Acknowledgements} \par %delete if not applicable))
This work was authored by the National Renewable Energy Laboratory, operated by Alliance for Sustainable Energy, LLC, for the U.S. Department of Energy (DOE) under Contract No. DE-AC36-08GO28308. Funding was provided by the U.S. Department of Energy's Office of Energy Efficiency and Renewable Energy (EERE) under the Solar Energy Technologies Office Award Number 38266. This report was prepared as an account of work sponsored by an agency of the United States Government. Neither the United States Government nor any agency thereof, nor any of their employees, makes any warranty, express or implied, or assumes any legal liability or responsibility for the accuracy, completeness, or usefulness of any information, apparatus, product, or process disclosed, or represents that its use would not infringe privately owned rights. Reference herein to any specific commercial product, process, or service by trade name, trademark, manufacturer, or otherwise does not necessarily constitute or imply its endorsement, recommendation, or favoring by the United States Government or any agency thereof. The views and opinions of authors expressed herein do not necessarily state or reflect those of the United States Government or any agency thereof. We would also like to thank the Institute for Solar Energy Research in Hamelin (ISFH) for providing the POLO IBC cells.

% References
\medskip

%\textbf{References}\\
\bibliographystyle{unsrt}
%\bibliography{TandemCore.bib, tempbib.bib}

\end{document}